\documentstyle[prl,aps,amsmath,amsfonts]{revtex}
\begin{document}
\newcommand{\ds}{\displaystyle}

\draft
\title{Anisotropy of Electrons Accelerated by a High-Intensity Laser Pulse}
\author{N.B. Narozhny\thanks{E-mail: narozhny@theor.mephi.ru}, M.S.
Fofanov\thanks{E-mail: fofanov@theor.mephi.ru}}
\address{Moscow State Engineering Physics Institute,
115409 Moscow, Russia}
\date{\today}
\maketitle

\begin{abstract}

We describe a realistic model for a focused high-intensity laser pulse in
three dimensions. Relativistic dynamics of an electron submitted to such
pulse is described by equations of motion with ponderomotive potential
depending on a single free parameter in the problem, which we refer to as
the "asymmetry parameter". It is shown that the asymmetry parameter can be
chosen to provide quantitative agreement of the developed theory with
experimental results of Malka~\textit {et al.} [Phys. Rev. Lett. {\bf 78},
3314 (1997)] who detected angular asymmetry in the spatial pattern of
electrons accelerated in vacuum by a high-intensity laser pulse.
\end{abstract}


In their recent paper, Malka~\textit {et al.}~\cite {MLM} have reported
experimental observation of electrons accelerated to relativistic energies
by a high-intensity linearly polarized subpicosecond laser pulse in vacuum
(see also comments on the paper ~\cite{MLM} and the author's reply in
Ref.~\cite {McD,MQ,MLM2}). This effect, known as high-intensity
ponderomotive scattering, was discussed in detail in Ref.~\cite {H}. It
occurs when the quiver amplitude imparted by the laser field to an electron
becomes comparable to the focal spot radius of the laser beam. If the beam
is Gaussian, the radial restoring force acting on the electron decays
exponentially and the electron can be scattered out of the pulse.

The data of Malka~\textit {et al.}~\cite {MLM} show that the energies
gained by the scattered electrons are in good quantitative agreement with
calculations of electron trajectories in the polarization plane made with
the first-order paraxial model for the laser field~\cite {MLM,MQ}.
Nevertheless, the first order paraxial model predicts isotropic electron
scattering~\cite {MQ} that is not supported by experimental results.
Indeed, accelerated electrons were detected by Malka~\textit {et al.} only
in the $\left({\bf E,k}\right)$~plane, while no significant signal was
detected after rotating the laser polarization direction by~$90^{o}$~\cite
{MLM}. In our opinion, this discrepancy between the theory and experiment
is due to the following. In the first-order paraxial model focusing of a
plane monochromatic wave only leads to an appearance of nonvanishing
longitudinal components of electromagnetic fields in the focal region.
However, the focusing is known to affect transverse components of the
fields also (see, e.g. Ref.~\cite {B}). In particular~\cite {B}, a plane
monochromatic wave, polarized linearly along the $x$~axis and propagating
along the $z$~axis, is converted by an aplantic system to a converging
spherical wave with non-vanishing $y$ and $x$~components of electric and
magnetic fields, respectively.

In this Letter, we show that experimentally observed~\cite{MLM} anisotropy
of ponderomotive electron scattering can be explained in the framework of a
realistic model for the laser field developed in our recent
paper~\cite{NF}. Our model is based on an exact solution of Maxwell
equations in three dimensions (3D), which can serve to describe a
stationary, focused monochromatic laser beam with characteristic frequency
$\omega$ and arbitrary intensity. Amplitudes of electric and magnetic
fields in the model depend on radial coordinates as well as the coordinate
along the direction of the beam propagation. These amplitudes are
characterized by parameters~$R$ and $L=\omega R^{2}$, which can be
interpreted as the focal spot radius and the Rayleigh length of the laser
beam, respectively. The model admits different field configurations, which
are determined by two coordinate functions satisfying certain second-order
partial differential equations. Some special choice of these functions
describes the Gaussian beams, which are widely used in optics. The model
can be generalized by introducing temporal amplitude envelope
$g(\varphi/\omega\tau)$, where $\varphi$ is the relativistically invariant
phase of the traveling wave, to describe a laser pulse with finite duration
$\tau$. (It is assumed that the function~$g(\varphi/\omega\tau)$ is equal
to unity at the point~$\varphi=0$ and decreases exponentially at the
periphery of the pulse for $|\varphi|\gg\omega\tau$.) In this case the
electric and magnetic fields of the model constitute an approximate
solution of Maxwell equations with the second-order accuracy with respect
to small parameters $\Delta$ and $\Delta^\prime$, defined as

\begin{equation}
\label{1} \Delta^\prime =1/\omega\tau\lesssim\Delta = 1/\omega R\ll 1.
\end{equation}

For a pulse propagating along the $z$~direction, $x$ and $y$~components of
the electric field oscillate with phase difference, which depends on $z$
and values of $x$ and $y$ coordinates of a point in the plain~$z=const$.
Moreover, the aforementioned phase difference depends also on $\varphi$.
Therefore, one cannot ascribe some definite type of polarization to a
tightly focused laser pulse. Nevertheless, for a weakly focused pulse
($\Delta \ll 1$), there always exists a region near the axis of the beam
$r\ll R$, where the field properties are very close to those of a plane
wave field. This region we call "the plane wave zone". It is reasonable to
ascribe polarization of the field in this region to the beam as a whole.
Hereafter we refer to the field of the pulse as linearly polarized in this
sense only. For a tightly focused beam the focal spot radius is of the
order of the wavelength and the plane wave zone doesn't exist. Therefore,
only polarization of the parental beam incident on the focusing optical
system can be ascribed to the focused beam in this case.

An arbitrary field, linearly polarized along the $x$~axis, may be
represented \cite{NF} as a superposition of $E$- and $H$-polarized waves
(i.e. waves with the vectors $\vec{E}$ and $\vec{H}$ being perpendicular to
the direction of the pulse propagation, compare with Ref.~\cite{BW}).
Relative contributions of $E$- and $H$-polarized waves to the resulting
field are characterized by the "asymmetry parameter"~$\mu$

\begin{equation}
\label{2} \mu =
\frac{E_{x0}^{h}-E_{x0}^{e}}{E_{x0}^{h}+E_{x0}^{e}},\quad\quad-\infty<\mu<\infty,
\end{equation}

\noindent where $E_{x0}^{e,h}$ are the $x$ components of the electric field
for $E$- and $H$-polarized waves at the focal point ${\bf r}=0$ for
$\varphi =0$. Note, that in contrast to the amplitude, the quantities
$E_{x0}^{e,h}$ can take both positive and negative values.

In the lowest approximation in $\Delta$ and $\Delta'$, the averaged
equations of motion of electrons (ponderomotive equations) in the field of
the linearly polarized laser pulse take the form \cite{NF}

\begin{equation}
\label{10}
\begin{array}{rclrcl} \ds{ \frac{d {\vec{q}_\perp}}{d\varphi}
}&=&\ds{ -\Delta\frac{m}{q_{-}}\frac{\partial U}{\partial
{\vec{\rho}_\perp}}}, &\qquad \ds{ \frac{d\vec{ \rho}_{\perp}}{d\varphi}}&
=&\ds{ \Delta\frac{\vec{q}_{\perp}}{q_{-}} },
\\{}\\ \ds{ \frac{d q_-}{d\varphi}}&=&0,
&\qquad \ds{ \frac{d \zeta}{d\varphi} }&=&\ds{ \Delta^2\frac{q_{z}}{q_{-}}
}.
\end{array}
\end{equation}

\noindent Here $q^{\mu}= \langle p^{\mu}\rangle$, where $p^{\mu}$~is
the~4-momentum of the electron, $q_{-}=q_{0}-q_{z}$, $\vec{\rho}_\perp=
\langle \vec{r}/R\rangle$, $\zeta=\langle z/R\rangle$, brackets
$\langle\rangle$ mean averaging over fast oscillations, and the
ponderomotive potential~$U$ is defined by the expression

\begin{equation}
\label{3} U=\frac{m\eta_0^2}{2}g^2(\varphi/\omega\tau) \left\{ \left|
F_1\right|^2 +\mu^2\left| F_2\right|^2 + \mu\cos 2\psi \left(
F_1F_2^*+F_1^*F_2\right)\right\},
\end{equation}

\noindent where $\tan\psi=\varrho_{x}/\varrho_{y}$,~ and $\eta_{0}$ is the
value of the dimensionless field intensity parameter

\begin{equation}
\label{4} \eta^{2}=\frac{e^{2}\langle
\textbf{E}^{2}\rangle}{m^{2}\omega^{2}},
\end{equation}

\noindent at the focal point at the moment $\varphi=0$
($\eta_0=a/\sqrt{2}$, where $a$ is the parameter of Malka~{\it et al.}).
Functions $F_{i}(\vec{\varrho}_{\perp},\zeta;\Delta)$~are chosen in the
form corresponding to the Gaussian beam \cite{NF}

\begin{equation}
\label{5}
\begin{array}{c} \ds{ F_{1} =
(1+2i\zeta)^{-2} \left\{1-\frac{{\varrho_{\perp}}^{2}}{1+2i\zeta}\right\}
\exp\left\{-\frac{{\varrho_{\perp}}^{2}}{1+2i\zeta}\right\}},
\\{}\\
\ds{F_{2}=-{\varrho_{\perp}}^{2}(1+2i\zeta)^{-3}\exp\left\{-\frac{{\varrho_{\perp}}
^{2}} {1+2i\zeta}\right\}}.
\end{array}
\end{equation}

The equation~(\ref{3}) shows that the ponderomotive potential $U$ depends
on the azimuthal angle $\psi$, and hence is generally speaking asymmetric.
The potential $U$ is symmetric only for the case $\mu=0$. The shape of the
ponderomotive potential in the plane~$z=0$ for $\varphi=0$ is shown in
Fig.\ref{fig:1} for the cases $\mu=0$ and $\mu=-1.55$.
Figure~\ref{fig:1}{\it a} represents the ponderomotive potential for the
standard case of Gaussian beam commonly used in literature, while
Figure~\ref{fig:1}{\it b} illustrates the dramatic difference between the
cases $\mu=0$ and $\mu\neq 0$. For $\mu\neq 0$, the ponderomotive
potential, possesses (besides the central peak) two extra maxima, which are
located in the polarization plane. They arise as a result of the
non-uniform intensity distribution in the plane~$z=0$ for~$\mu\neq 0$.
Locations of the additional maxima, as well as their amplitudes, are
determined by the value of $\mu$. It is noteworthy, that the case $\mu=0$
is the only one when  $E_{y}$- and $H_{x}$-components of the electric and
magnetic fields remain to be equal to zero outside the plane wave zone for
the pulse polarized along the $x$~axis.

We use Eqs. (\ref{10}) and (\ref{3}) for our analysis of free electron
acceleration by a co-propagating intense laser pulse in vacuum under
conditions close to those used in the experiments of Malka~{\it et
al.}~\cite{MLM}. The initial electron energy is taken to be
$\varepsilon=10$~keV ($v_0=0.2c$). The laser field parameters are:
$\lambda=1$~$\mu$m, $\eta_0=2.12$ (corresponds to the parameter $a=3$ of
Malka~{\it et al.}), $R=10$~$\mu$m and $\omega\tau=480$. For the temporal
envelope of the pulse $ g(\varphi/\omega\tau)$ a sine-squared shape is
taken. The asymmetry parameter is not determined experimentally and remains
a free parameter of the problem. Its value, $\mu=-1.55$, has been chosen
for better fitting of our computational results to the results of the
experiment.

It is clear that the maximum energy will be gained by electrons that
initially propagate exactly along the axis of the laser beam. However, the
ponderomotive equations yield zero net energy transfer for such electrons,
since they can "feel" the spatial gradient of the Gaussian laser field, and
hence can be scattered out of the pulse, only due to the quiver motion
which is absent for the average trajectory described by the ponderomotive
equations. Nevertheless, there exist a family of trajectories with nonzero
initial distances from the beam axis, for which the gained energy is close
to its maximum value. The energy and scattering angle of the electrons
depend also on the position at which the particle is overtaken by the pulse
\cite{MLM}. Indeed, the maximum energy is obtained for the electrons that
experience the peak field of the laser and therefore meet the laser pulse
at some distance before the focus. As a result, we obtain a 3D domain of
injection positions of electrons obtaining the final kinetic energy~$W\geq
0.9$~MeV at the scattering angle $39.5^{o}$.

Different cross-sections of this domain for the potential with $\mu=-1.55$
(Fig.~\ref{fig:1}{\it b}) are shown in Fig.~\ref{fig:2}. The longitudinal
size of the domain is of the order of the Rayleigh length for the laser
beam~$L$, whereas its transverse size is much less than the focal spot
radius~$R$. The cross sections in Fig.~\ref{fig:2} display high degree of
radial anisotropy. Their shape, of course, is essentially determined by the
type of the ponderomotive potential or by the value of $\mu$. In
particular, the domain of injection positions for the case (not presented
here for the sake of compactness) of the potential shown in
Fig.~\ref{fig:1}{\it a} with $\mu=0$ is purely radial, in agreement with
\cite{MQ}.

To obtain the angular distribution of scattered electrons we have
calculated their trajectories numerically, applying the fourth-order
Runge-Kutta method to the ponderomotive equations (\ref{10}). Initial
positions of the electrons were taken from the domain shown in
Fig.~\ref{fig:2}. Since the shape of the cross-sections $z=const$ varies
very slowly at intervals $\delta z \sim R$, we have considered equidistant
planes~$z=nR$ with $n=-27,-26,\ldots,5$. In each of these cross-sections,
the electron injection positions were chosen randomly under condition that
their density was constant and equal to $3\times 10^{15}/R^2$. Physically,
such procedure corresponds to uniformity of the initial electron beam. The
total number of electron trajectories considered in such a way was more
than $2\times 10^{7}$. We were interested only in those trajectories, which
crossed the plane $z=11.66$~cm at the points contained inside the ring with
radii $r_{1}=8.99$~cm and $r_{2}=9.89$~cm. The latter conditions were
determined by the position and angular size of the detector in the
experiment \cite{MLM}.

The results of the calculations are presented in Fig.~\ref{fig:3}. We plot
the normalized number of scattered electrons $\langle n\rangle$ with final
energies~$W\geq 0.9$~MeV as a function of the azimuthal scattering
angle~$\alpha$. The values of $\langle n\rangle$ at any given~$\alpha$ were
obtained by averaging the number of scattered electrons over the range of
$\alpha$ equal to the angular size of the detector used in the
experiments~\cite{MLM}. It is easily seen, that the angular distribution of
scattered electrons essentially depends on the parameter of asymmetry
$\mu$. At $\mu=0$, which corresponds to the symmetrical Gaussian
ponderomotive potential shown in Fig.~\ref{fig:1}{\it a}, the distribution
is isotropic and purely radial (compare to the result of Ref.~\cite{MQ}).
At the same time, if $\mu=-1.55$, the number of scattered electrons
detected in the $\left({\bf E,k}\right)$~plane ($\alpha=0$) is about 30
times higher than that in the $\left({\bf H,k}\right)$~plane
($\alpha=\pi/2$). This result is in good quantitative agreement with
observations of Malka~\textit {et al.} \cite{MLM}.

The 30-fold anisotropy of accelerated electrons is clearly explained by
asymmetry of the ponderomotive potential (\ref{3}). The cross term in
(\ref{3}), besides asymmetric corrections to the radial force, gives rise
to a tangential force which is responsible for pushing electrons out of the
plain perpendicular to the polarization plane. The latter corresponds to
minimum of the ponderomotive potential as function of azimuthal angle
$\psi$, while the perpendicular plane to its maximum. Therefore one could
be surprised that the number of electrons scattered at the angle
$\alpha=\pi/2$ is not equal to zero. Certainly it is explained by
complicated structure of the ponderomotive potential (\ref{3}), namely by
the fact that the cross term can change its sign at the periphery of the
focus.

The asymmetry of the ponderomotive potential itself is determined by
non-zero value of the parameter $\mu$ which characterizes relative
contributions of E- and H-polarized waves to the resulting field. As far as
we know, nobody has never controlled the parameter $\mu$ in experiments.
The reason is evident. Before the work of Malka {\it et al} \cite{MLM}
there were no experiments where the three-dimensional intensity
distribution of a laser pulse influenced physical results. Therefore even
qualitative coincidence of our calculations with the results of the
experiment \cite{MLM} would give a deeper insight into physics of the
electron-laser interaction. However, it appeared that the model describes
the experiment quantitatively. Of course, the quantitative agreement
between the developed theory and the experiment is based on fitting of a
single free parameter $\mu$ in the problem, which has not been measured
experimentally. Therefore, from the standpoint of our model, the experiment
of Malka~\textit {et al.} \cite{MLM} could be considered as a probe for 3D
field distribution in the laser pulse. The correctness of our approach
could be verified by an independent experiment for another physical
situation performed with the same laser system. Measurements of angular
distribution of ATI electrons could serve as a good example of such
experiment.

We thank M.V. Fedorov and V.D. Mur for fruitful discussions. This work was
supported by the Russian Foundation for Basic Research under projects
00-02-16354 and 00-02-17078.

\begin{figure}
 \caption{Ponderomotive potential~(\ref{3}) in the plane~$z=0$ at $\varphi=0$
 for two values of asymmetry parameter:(a) $\mu=0$ and (b) $\mu=-1.55$.}
 \label{fig:1}
\end{figure}
\begin{figure}
 \caption{Cross sections of 3D injection-position domain for electrons gaining
 the energy~$W\geq 0.9$~MeV in the ponderomotive potential~(\ref{3}) with $\mu=-1.55$.
 $z$~coordinates  of the cross sections are (a) $z=-21R$,
 (b) $z=-16R$, (c) $z=-11R$, (d) $z=-6R$, (e) $z=-R$, and (f) $z=2R$.}
 \label{fig:2}
\end{figure}
\begin{figure}
 \caption{ Normalized number of scattered electrons with the energy~$W\geq 0.9$~MeV
 as a function of azimuthal angle $\alpha$ for (a) $\mu=0$ and (b) $\mu=-1.55$.}
 \label{fig:3}
\end{figure}

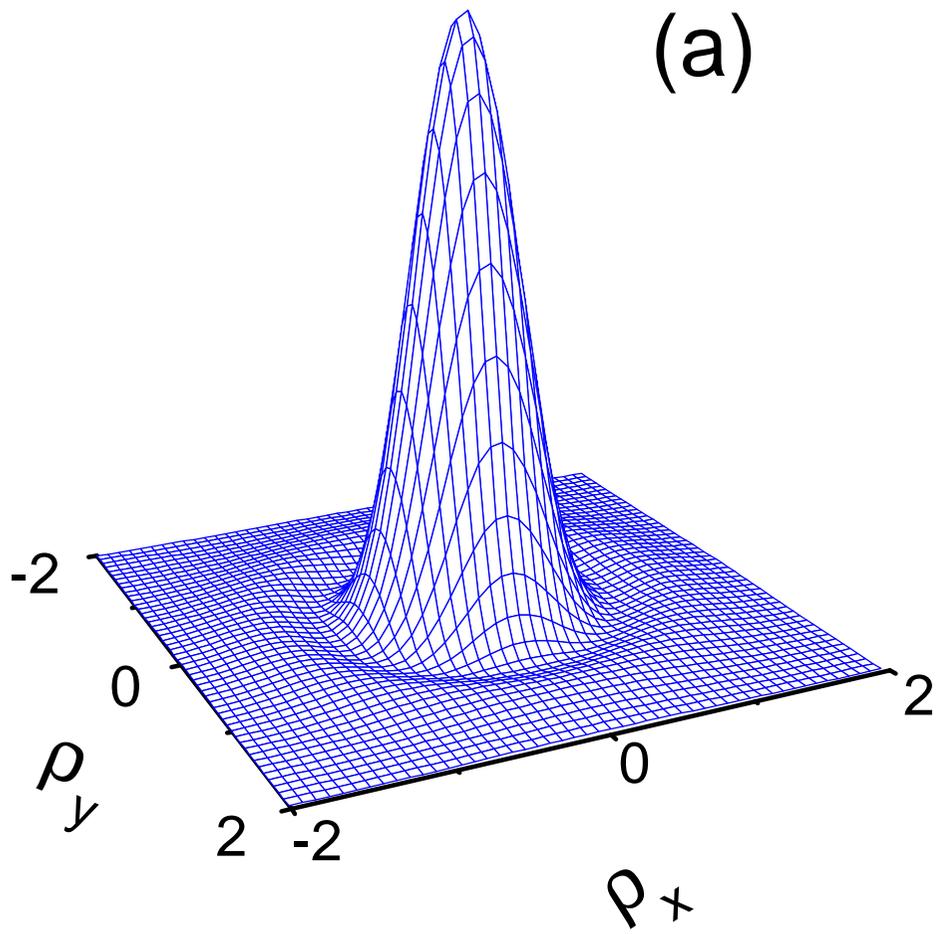
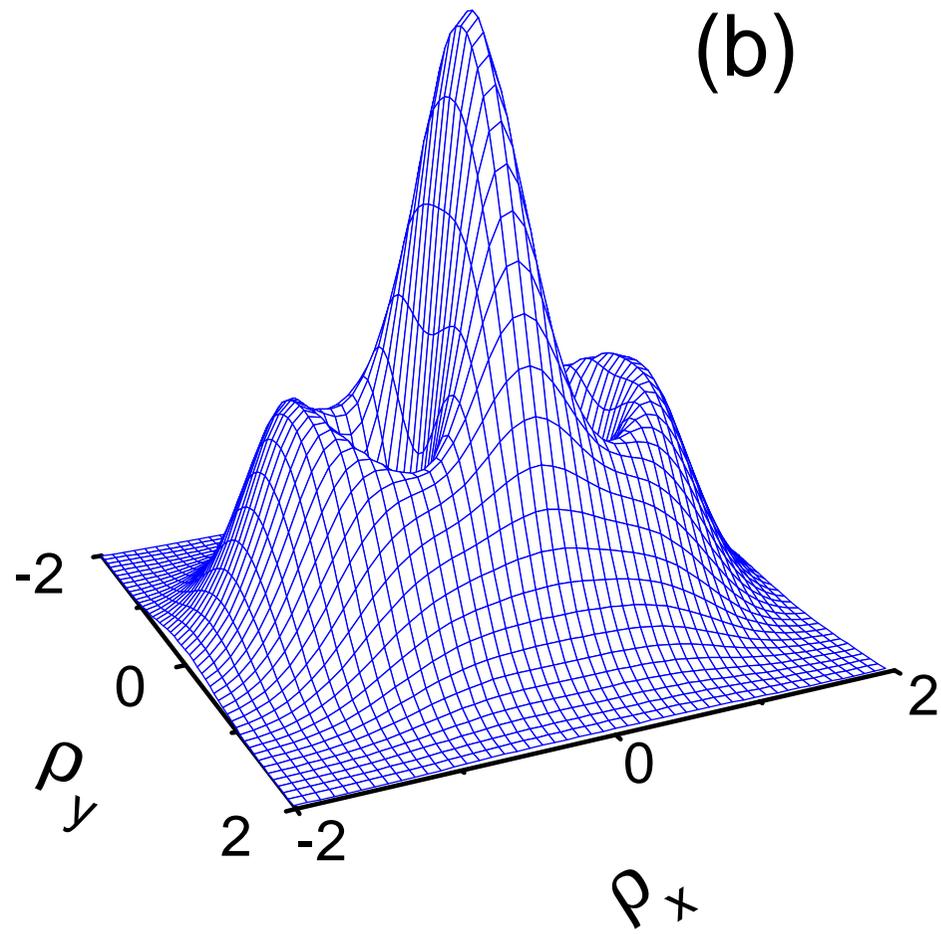
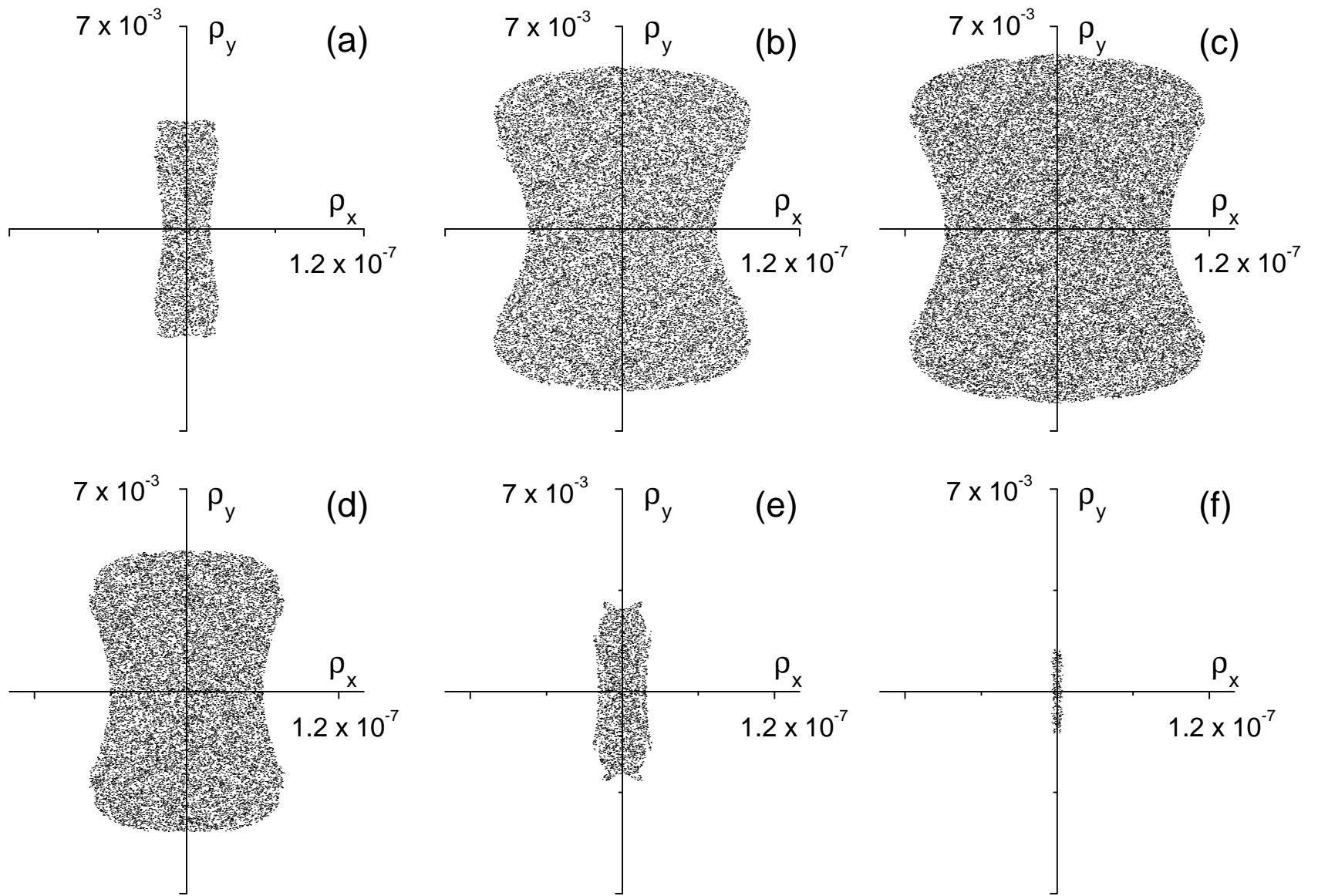
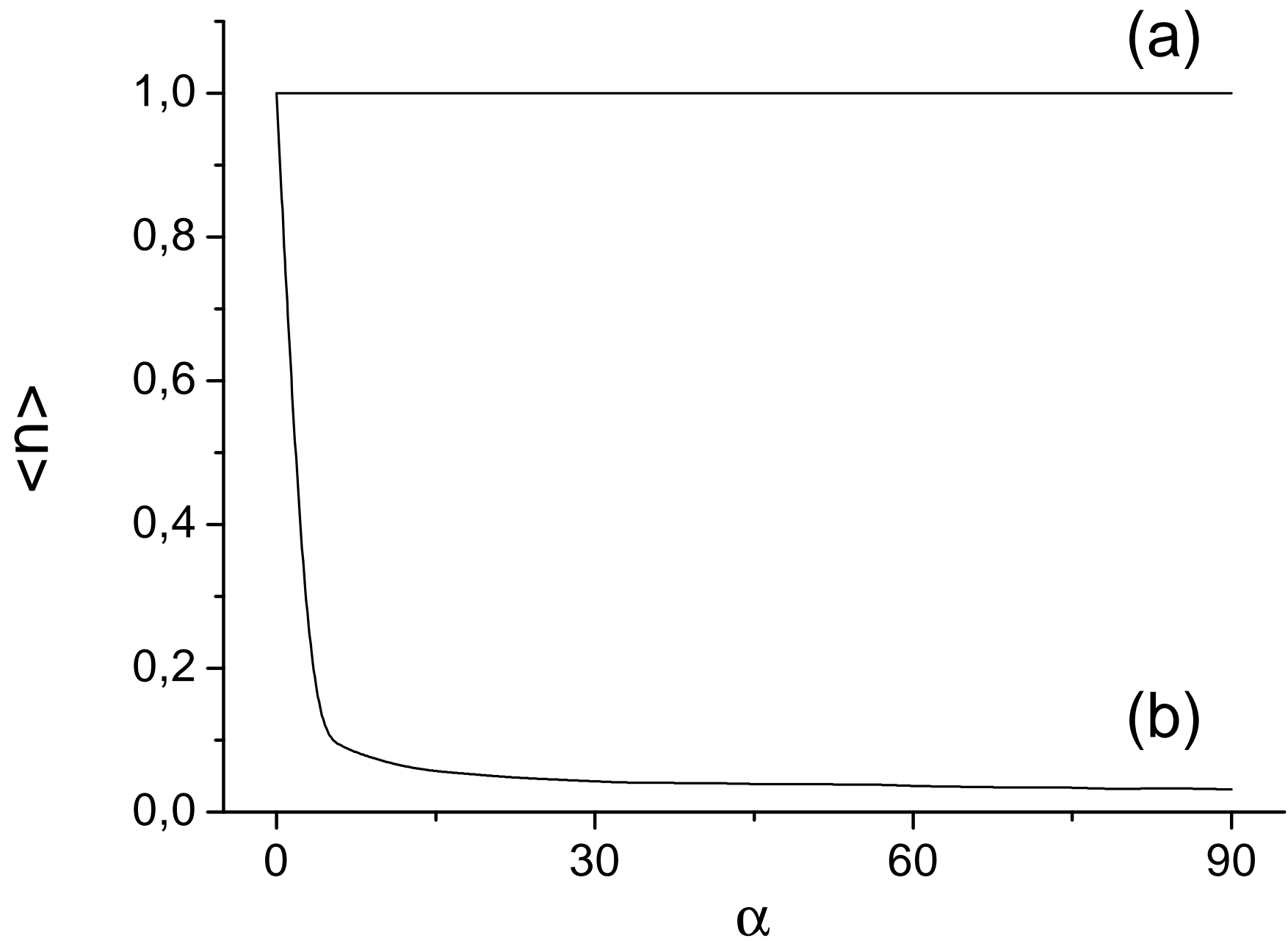
\end{document}